\documentclass[aps,prl,twocolumn,showpacs,superscriptaddress]{revtex4-2}
\usepackage{amsmath,amssymb}
\usepackage{graphicx}
\graphicspath{{./figures/}}
\usepackage{wasysym}
\usepackage{amsfonts}
\usepackage{bm}
\usepackage{enumerate}
\usepackage{color}
\usepackage{url}

\usepackage{hyperref}
\hypersetup{
	colorlinks = true,
	urlcolor = blue,
	linkcolor = blue,
	citecolor = blue
}

\usepackage{epstopdf}
\usepackage{latexsym}
\usepackage[caption=false,singlelinecheck=false]{subfig}

\newcommand{\bra}[1]{\ensuremath{\langle #1 |}}
\newcommand{\ket}[1]{\ensuremath{| #1 \rangle}}

\newcommand{\bea}{\begin{eqnarray}}
\newcommand{\eea}{\end{eqnarray}}
\newcommand{\be}{\begin{eqnarray}}
\newcommand{\ee}{\end{eqnarray}}
\newcommand{\bw}{\begin{widetext}}
\newcommand{\ew}{\end{widetext}}

\begin{document}

\title{Nonlinear spectroscopy of bound states in perturbed Ising spin chains}
	
\author{GiBaik Sim}
\affiliation {Department of Physics TQM, Technische Universit\"{a}t M\"{u}nchen, $\&$ James-Franck-Straße 1, D-85748 Garching, Germany}
\affiliation{Munich Center for Quantum Science and Technology (MCQST), 80799 Munich, Germany}

\author{Johannes Knolle}
\affiliation {Department of Physics TQM, Technische Universit\"{a}t M\"{u}nchen, $\&$ James-Franck-Straße 1, D-85748 Garching, Germany}
\affiliation{Munich Center for Quantum Science and Technology (MCQST), 80799 Munich, Germany}
\affiliation{Blackett Laboratory, Imperial College London, London SW7 2AZ, United Kingdom}

\author{Frank Pollmann}
\affiliation {Department of Physics TQM, Technische Universit\"{a}t M\"{u}nchen, $\&$ James-Franck-Straße 1, D-85748 Garching, Germany}
\affiliation{Munich Center for Quantum Science and Technology (MCQST), 80799 Munich, Germany}

\date{\today}

\begin{abstract}
	We study the nonlinear response of non-integrable 1D spin models using infinite matrix-product state techniques. As a benchmark and demonstration of the method, we first calculate the 2D coherent spectroscopy for the exactly soluble ferromagnetic transverse field Ising model, where excitations are freely moving domain-walls. We then investigate the distinct signatures of confined bound states by introducing a longitudinal field and observe the emergence of strong non-rephasing like signals. To interpret the observed  phenomena, we use a two-kink approximation to perturbatively compute the 2D spectra. We find good agreement in comparison with the exact results of the infinite matrix-product state method in the strongly confined regime. We discuss the relevance of our results for quasi$-$1D Ising spin chain materials such as $\mathrm{CoNb}_2\mathrm{O}_6$.
\end{abstract}
	
\maketitle
{\em Introduction.}
Spectroscopic tools have played a crucial role in our understanding of complex quantum systems~\cite{devereaux2007inelastic}. However, we can only partially measure their correlations with existing tools. Terahertz 2D coherent spectroscopy,  one of developing spectroscopic tools~~\cite{shen1984principles, hamm2011concepts,mukamel1999principles}, stands out as a technique for a deeper understanding of strongly-correlated condensed matter systems. It probes the nonlinear optical response which has been used to identify the ground state symmetry of magnetic~\cite{fiebig2005second} and superconducting~\cite{chu2020phase,schwarz2020classification} materials, geometric phase in topological materials~\cite{wu2017giant,shao2021nonlinear,he2021quantum}, novel ground states in correlated systems~\cite{zhao2017global,zhao2016evidence}, and quasiparticle decay processes in a disordered system~\cite{mahmood2021observation}. Besides, recent experimental advances with terahertz sources put the technique in a proper energy range to study rotational dynamics in molecules~\cite{lu2016nonlinear}, spin waves in conventional magnets~\cite{lu2016nonlinear}, and exotic excitations in quantum magnets~\cite{wan2019resolving,li2021photon,choi2020theory,nandkishore2021spectroscopic,fava2021hydrodynamic}.
	
In contrast to more common 1D spectroscopy, the 2D extension unravels not only the optical excitations but also their interplay~\cite{mukamel1999principles,hamm2011concepts}. The advantage of this experimental technique has been widely adopted by chemists to reveal the structure of complex molecules with great success. However, such achievements rely on powerful numerical methods that help to interpret complicated experimental data starting from concrete microscopic models~\cite{woutersen2002peptide,cho2008coherent,terranova2014molecular}. In this regard, it is desirable to develop an efficient numerical platform for future 2D spectroscopy experiments on quantum magnets similar to the successful use of matrix product state (MPS) techniques for conventional 1D spectroscopy~\cite{paeckel2019time,vanderstraeten2015scattering,bera2017spinon,vanderstraeten2018quasiparticles,van2021efficient,fava2020glide,morris2021duality}. However, the calculation of nonlinear response is less explored and the need for multiple time evolutions makes it much more challenging. 

In this work, we propose an efficient numerical tool using infinite MPS (iMPS) and study the nonlinear response of 1D spin model. To benchmark the method, we first focus on the 1D transverse field Ising model (TFIM) whose nonlinear response can be analytically calculated using Jordan-Wigner (JW) transformations~\cite{wan2019resolving}. Motivated by the quasi-1D structure of $\mathrm{CoNb}_2\mathrm{O}_6$ -- one of the best material example of an Ising chain magnet -- (albeit with more complicated magnetic interactions~\cite{fava2020glide,morris2021duality}) we then include longitudinal field terms which capture the  effects of inter-chain interactions and lead to the emergence of confined bound state excitations~\cite{lee2010interplay,kinross2014evolution,xu2022quantum}. As a consequence, new signals appear in 2D spectroscopy which include strong non-rephasing and rephasing like peaks. Our results from the iMPS method are furthermore corroborated by perturbative calculations starting from the projected two-kink (TK) low energy subspace. We find quantitative agreement in the strongly confined regime, which allows us to understand the origin of sharp peaks in the 2D spectrum as transitions between bound states.  


{\em Model.}	
We first introduce the 1D TFIM 
\bea
H_0=-J\sum_n \sigma_n^z \sigma_{n+1}^z-h^x\sum_n \sigma_n^x 
\label{eq:tfim_l}
\eea
with $J,h_x>0$. For $h_x<h^c_x=J$, it stabilizes a doubly degenerate ferromagnetic ground state polarized along the easy axis $\hat z$. When $h_x>h^c_x$, the system has a unique paramagnetic ground state. In the ferromagnetic regime, the experimental excitation, i.e., a local spin flip, splits into two freely moving kinks (domain-walls) between two degenerate states. In the context of non-linear spectroscopy these fractionalized excitations have been shown to be manifest as sharp signatures in the third order magnetic susceptibilities~\cite{wan2019resolving}.

We now include a longitudinal field and focus on the Hamiltonian given by
$H=H_0 - h^z\sum_n \sigma^z_n$ with $J,h_x,h_z>0$. In the  ferromagnetic regime, the longitudinal field lifts the degeneracy and selects one of the polarized ground states. Besides, it induces a linear confining potential between the kinks leading to bound states. As a result, the broad continuum of free kink excitations as probed in linear response fragments into sharp peaks~\cite{kjall2011bound}.  At a low transverse field, the splitting can be understood via a Schr\"{o}dinger equation for the relative kink separation with a linear potential~\cite{mccoy1978two} which can also be generalized to include lattice effects~\cite{coldea2010quantum,morris2014hierarchy,rutkevich2010weak,rutkevich2008energy,shinkevich2012spectral,kormos2017real,liu2019confined}. The main questions of our work are: How to efficiently simulate the nonlinear response of the TFIM with a longitudinal field using MPS methods? What are the robust signatures of confined bound states in nonlinear 2D spectroscopy?
	

{\em 2D spectroscopy.}
Here, we introduce a two-pulse protocol which following previous work Ref.~\onlinecite{wan2019resolving}. In this setup, two Dirac-delta pulses $B_0$ and $B_{\tau}$ which are polarized along $\hat \alpha$ and $\hat \beta$ directions, respectively, reach the sample at time $T=0$ and $T=\tau>0$ successively. These magnetic pulses couple to the local moments of the sample and the induced magnetization along $\hat \gamma$ direction is recorded as $M^\gamma_\mathrm{0\tau}(T)$ at time $T = \tau + t$ where $t>0$ is the time interval between the second pulse $B_\tau$ and the measurement. To subtract the signal from the linear response, two different experiments are repeated but with pulse $B_0$ or $B_{\tau}$ alone to measure $M^\gamma_0(T)$ and $M^\gamma_\tau(T)$. The nonlinear signal field emerging from the sample at $T = \tau + t$ in the $\hat \gamma$ direction is defined as
\bea
M^\gamma_{NL}(T) \!\equiv\! M^\gamma_\mathrm{0\tau}(T) - M^\gamma_0(T) - M^\gamma_\mathrm{\tau}(T).
\label{eq:signal_define}
\eea
The nonlinear signal depends only on the nonlinear responses and directly measures the second and higher order magnetic susceptibilities~\cite{nandkishore2021spectroscopic}:
\bea
M^\gamma_\mathrm{NL}(t, \tau) &=& B_0 B_\tau \chi_{\gamma\beta\alpha}^{(2)}(t,\tau + t ) \nonumber \\
&+& (B_0)^2 B_\tau \chi_{\gamma\beta\alpha\alpha}^{(3)}(t,\tau + t ,\tau + t ) \nonumber \\ 
&+& B_0 (B_\tau)^2 \chi_{\gamma\beta\beta\alpha}^{(3)}(t,t,\tau + t )+O(B^4).
\label{eq:signal}
\eea
	
The 2D spectrum is the Fourier transform of $M^\gamma_\mathrm{NL}(t, \tau)$ over both time domains $t$ and $\tau$. In Eq.~(\ref{eq:signal}), the leading  contribution to the nonlinear response in the two-pulse setup, i.e., the second order nonlinear susceptibility $\chi_{\gamma\beta\alpha}^{(2)}(t,\tau + t )$, is given as
\bea
\nonumber
\chi^{(2)}_{\gamma\beta\alpha}(t,\tau+t) = &-& \frac{\theta(t)\theta(\tau)}{4L} \mathrm{Re}\sum_{j,l,m}\big[ S^{\gamma\beta\alpha}_{j,l,m}(\tau+t,\tau,0)\\ 
&-& S^{\beta\gamma\alpha}_{j,l,m}(\tau,\tau+t,0) \big]	
\label{eq:chi2}
\eea
with the three point spin correlation function in the ground state $\ket \psi$,
\bea
S^{\gamma\beta\alpha}_{j,l,m}(T_1,T_2,0) = \bra \psi \sigma^\gamma_j(T_1) \sigma^\beta_l(T_2) \sigma^\alpha_m (0) \ket \psi
\label{eq:t_point}
\eea
where $\sigma^\gamma_j(T) \equiv e^{iHT} \sigma^\gamma_j e^{-iHT}$. When the Hamiltonian and $\ket \psi$ preserves the lattice translation symmetry, the site index $m$ in Eq.~(\ref{eq:chi2}) and Eq.~(\ref{eq:t_point}) can be fixed, e.g. as $c \equiv L/2$ the central site of the system, which  we use in the following. Then, $\chi^{(2)}_{\gamma\beta\alpha}(t,\tau+t)$ is obtained by evaluating
\bea
\nonumber
- \frac{\theta(t)\theta(\tau)}{4} \mathrm{Re}\sum_{j,l}\big[ S^{\gamma\beta\alpha}_{j,l,c}(\tau+t,\tau,0)- S^{\beta\gamma\alpha}_{j,l,c}(\tau,\tau+t,0) \big].
\\
\label{eq:chi2_t}
\eea


{\em Method.}
A promising tool for calculating $S^{\gamma\beta\alpha}_{j,l,c}(\tau+t,\tau,0)$ in Eq.(\ref{eq:chi2_t}) for a whole range of site indices $j$ and $l$ is to use the iMPS method. We only need to perform two different real time evolution runs to calculate $\sum_{j,l}S^{\gamma\beta\alpha}_{j,l,c}(\tau+t,\tau,0)\!=\!\sum_{j,l} e^{iE(\tau+t)}\bra \psi \sigma^\gamma_j e^{-iHt} \sigma^\beta_l e^{-iH\tau} \sigma^\alpha_{c} \ket \psi$ where $E$ is the ground state energy. In addition,  finite size effects are avoided. Such effect originates from the bouncing of correlations following a local quench at site $j=1$ or $L$, boundary sites of the system. Below, we explain a procedure to obtain $\bra \psi \sigma^\gamma_j e^{-iHt} \sigma^\beta_l e^{-iH\tau} \sigma^\alpha_{c} \ket \psi$ (See Section I of Supplementary Information (SI) for $S^{\beta\gamma\alpha}_{j,l,c}(\tau,\tau+t,0)$).

\begin{figure}[t!]
	\centering
	\includegraphics[scale=0.53]{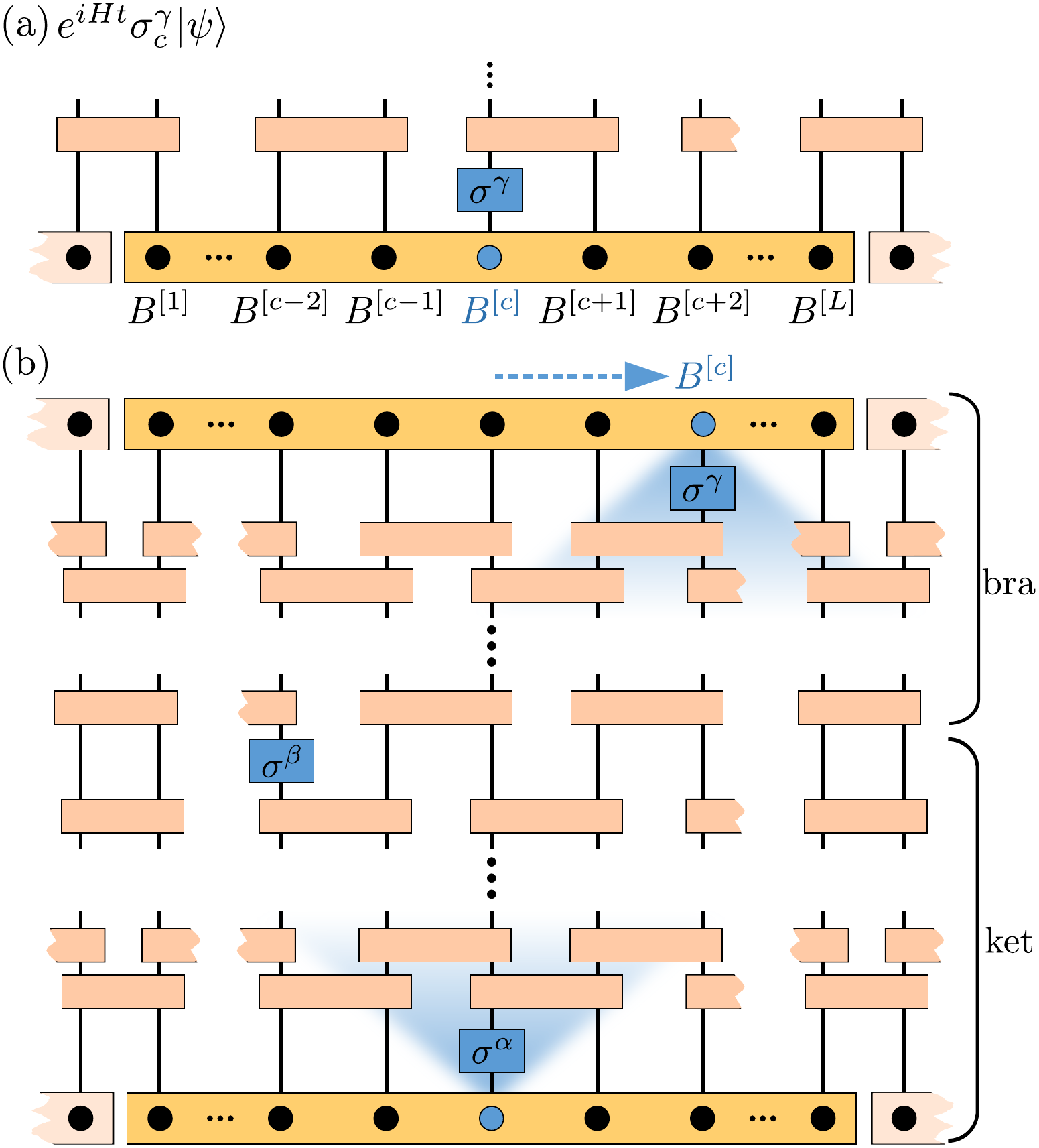}
	\caption{(color online) Exploiting infinite boundary conditions for translation-invariant systems, two real time evolution runs are sufficient to evaluate $\sum_{j,l}S^{\gamma\beta\alpha}_{j,l,c}(\tau+t,\tau,0)$. (a) iMPS representation of a state $e^{iHt}\sigma^\gamma_{c}\ket{\psi}$ with L sites in the unit cell. (b) Transfer matrix of two distinct iMPS ``bra" and ``ket" which represent $e^{iHt}\sigma^\gamma_{c+2}\ket{\psi}$ and $\sigma^\beta_{c-2} e^{-iH\tau} \sigma^\alpha_{c} \ket \psi$ respectively. The color gradient illustrates the light cone spreading of correlations following a local quench.} 
	\label{fig:method}
\end{figure}
	\begin{enumerate}
		\item
		Find a ground state and perform a time evolution following a local quench, $\sigma_c^{\gamma}$ or $\sigma_c^{\alpha}$, using infinite time evolving block decimation (iTEBD) method\cite{vidal2003efficient,vidal2004efficient,vidal2007classical} to obtain an iMPS for $e^{iHt} \sigma^\gamma_c \ket \psi$, which is shown in Fig.~\ref{fig:method}(a), or $e^{-iH\tau} \sigma^\alpha_{c} \ket \psi$.
		\item
		Shift every $B$ tensor of an iMPS, which represents $e^{iHt} \sigma^\gamma_c \ket \psi$, $(c-j)$ sites to the right within a window of size $L$ to obtain a new iMPS ``bra" associated to $e^{iHt}\sigma^\gamma_{j}\ket{\psi}$.
		\item
		Apply a local operator $\sigma^\beta_{l}$ to an iMPS associated to $e^{-iH\tau} \sigma^\alpha_{c} \ket \psi$ and get a new iMPS ``ket" which represents $\sigma^\beta_l e^{-iH\tau} \sigma^\alpha_{c} \ket \psi$.
		\item
		Evaluate an overlap of two iMPS ``bra" and ``ket" within the window by calculating the dominant left and right eigenvector of the corresponding transfer matrix, which is shown in Fig.~\ref{fig:method}(b), and obtain $\bra \psi \sigma^\gamma_j e^{-iHt} \sigma^\beta_l e^{-iH\tau} \sigma^\alpha_{c} \ket \psi$~\cite{kjall2011bound}.
	\end{enumerate}  

\begin{figure}
	\centering
	\includegraphics[scale=0.8]{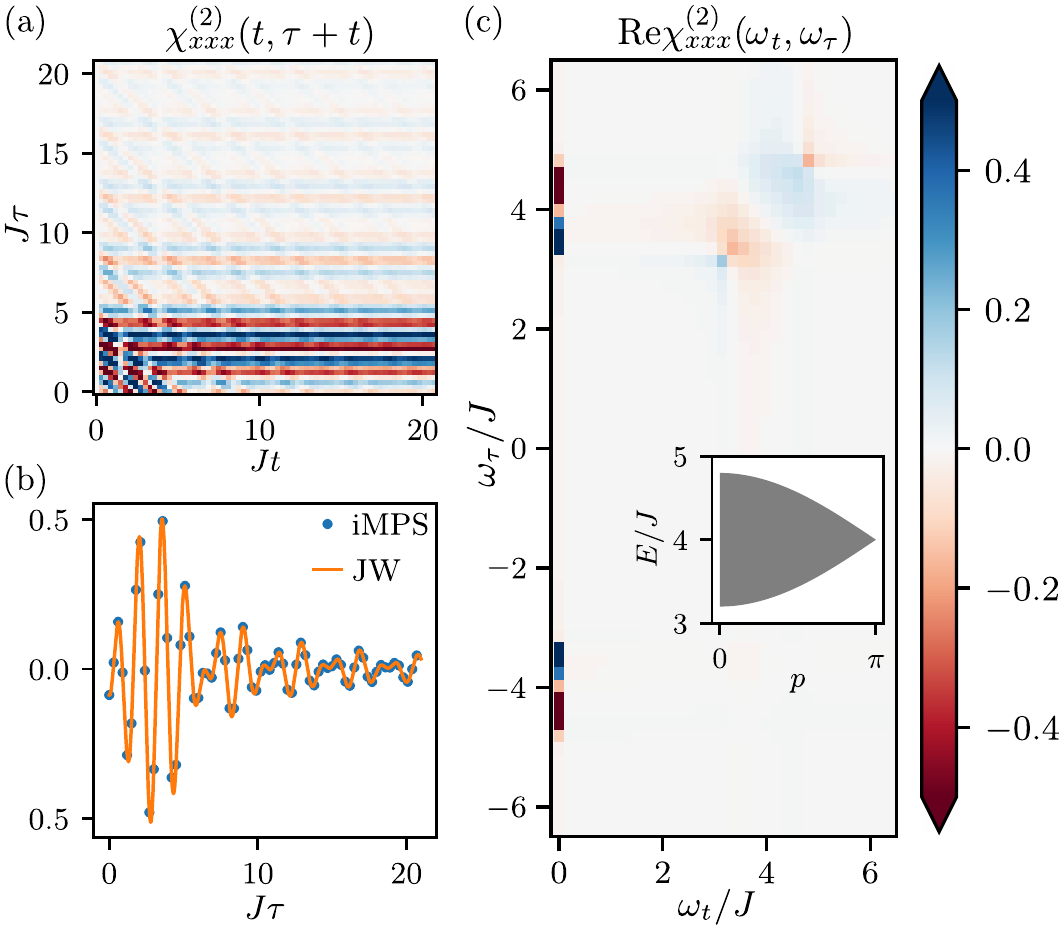}
	\caption{(color online) Second order susceptibility in the ferromagnetic phase of the TFIM with $h_x/J=0.2$. (a) $\chi^{(2)}_{xxx}(t, \tau+t)$ from iMPS method with a window of size $L=80$. The data is rescaled such that the maximal absolute value is 1. (b) $\chi^{(2)}_{xxx}(t, \tau+t)$ at $Jt=15$ from iMPS method and JW formalism. For the latter case, we set a size of the system $L=80$ with PBC. (c) Real part of Fourier transformed $\chi^{(2)}_{xxx}(t,\tau+t)$. Inset: TK excitation continua of the TFIM.}
	\label{fig:tfim}
\end{figure}

Before investigating the 2D spectrum of the non-integrable TFIM with longitudinal field, we first focus on the free TFIM and compare the result of $\chi^{(2)}_{xxx}(t,\tau+t)$ using two different schemes, i.e., the numerical iMPS method and analytic calculations via the JW transformation with periodic boundary condition (PBC). Here and below, we set $h_x/J=0.2$ and all iMPS simulations are done with a spatial window of size $L=80$ sites and over the time range $Jt,J\tau=30$. Within such temporal range, the light cone spreading of correlations, which follows a local quench at the center of a spatial window, does not reach the boundary of the window. In Fig.~\ref{fig:tfim}(a), we plot the result of $\chi^{(2)}_{xxx}(t,\tau+t)$ from the iMPS method. In order to check the errors of our method, we tracked the truncation error, the truncated weight of many-body wave function at each time step in iTEBD, which quantifies an upper limit for the truncation effect on local observables ($ \lesssim 10^{-8}$ for every result given in our study). Besides, we also followed the dependence of $\chi^{(2)}_{xxx}(t,\tau+t)$ on the time step $\delta t$ and the bond dimension $\chi$, fixing to $\delta t=0.03 / J$ and $\chi_{\text{max}}=30$. In Fig.~\ref{fig:tfim}(b), we compare $\chi^{(2)}_{xxx}(t,\tau+t)$ at $Jt=15$ from the iMPS method with the one from the JW formalism which confirms exact agreement. In Fig.~\ref{fig:tfim}(c), we plot Re$\chi_{xxx}^{(2)}(\omega_t, \omega_\tau)$, the real part of the Fourier transformed $\chi^{(2)}_{xxx}(t,\tau+t)$. It contains a sharp vertical line of intensity centered at $\omega_t=0$. Regarding $\omega_t$ and $\omega_\tau$ as the detecting and pumping frequencies, the response is known as a rectification signal. It also contains a diffusive, weak non-rephasing signal in the first frequency quadrant, mirroring the energy range of the free TK continuum, which is shown in the inset of Fig.~\ref{fig:tfim}(c)~\cite{wan2019resolving}.

\begin{figure}
	\includegraphics[scale=0.85]{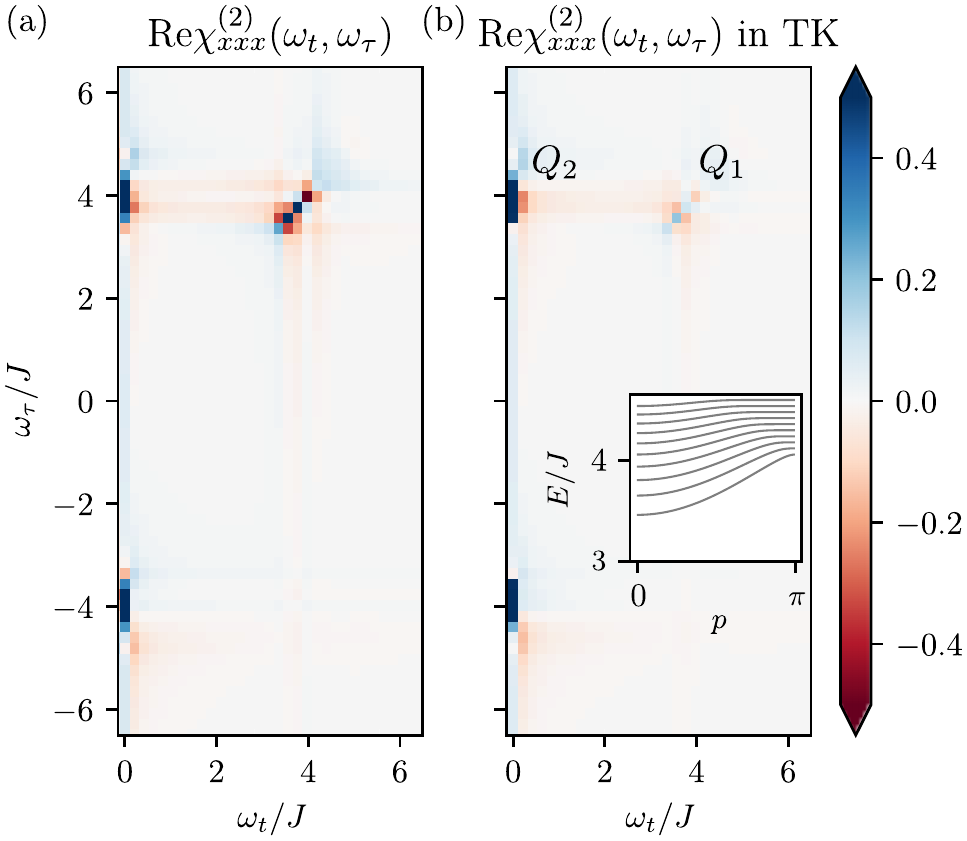}
	\caption{(color online) The 2D spectra $\chi^{(2)}_{xxx}(\omega_t, \omega_\tau)$ of TFIM with longitudinal field for $h_x/J = 0.2$ and $h_z/J = 0.03$. For the spectra, first and fourth quadrant are only shown. The other half are obtained by complex conjugation.(a) Result of iMPS method. (b) Result of perturbative calculation within projected TK subspace. Inset : The energy spectrum of TK bound states in the weakly confined regime.} 
	\label{fig:hz_1}
\end{figure}
\begin{figure}
	\includegraphics[scale=0.85]{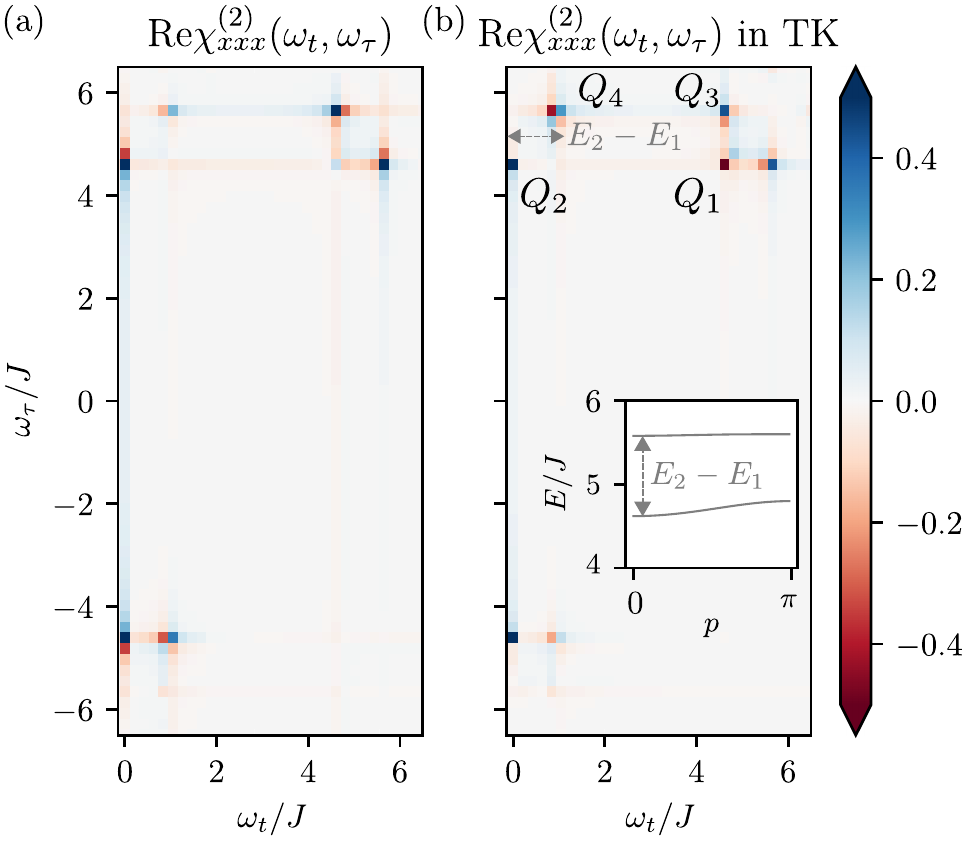}
	\caption{(color online) 2D spectra $\chi^{(2)}_{xxx}(\omega_t, \omega_\tau)$ of TFIM with longitudinal field for $h_x/J = 0.2$ and $h_z/J = 0.4$ (a) Result of iMPS method. (b) Result of perturbative method. Inset: The energy spectrum of TK bound states in the strongly confined regime.}
	\label{fig:hz_2}
\end{figure}

{\em Results.}	
Next, we focus on the 2D spectrum of the 1D TFIM with longitudinal field using the iMPS method. Fig.~\ref{fig:hz_1}(a) shows the Re$\chi_{xxx}^{(2)}(\omega_t, \omega_\tau)$, which is calculated in weakly confined regime with $h_z/J=0.03$ (See Section II of SI for the Im$\chi_{xxx}^{(2)}(\omega_t, \omega_\tau)$, which is related to the Re$\chi_{xxx}^{(2)}(\omega_t, \omega_\tau)$ by the dispersion relation\cite{kogan1963electrodynamics,caspers1964dispersion,bassani1991dispersion}). This value is similar to the one used in Ref.~\onlinecite{coldea2010quantum} to describe $\mathrm{CoNb}_2\mathrm{O}_6$. New spectroscopic signals are encoded in $\chi_{xxx}^{(2)}(\omega_t, \omega_\tau)$ in the presence of a longitudinal field. First, it contains a dominant non-rephasing signal which appears as diagonal peaks in the first quadrant. At the same time, a weakly diffusive terahertz rectification signal is also detected as a streak along the $\omega_\tau$ axis. In Fig.~\ref{fig:hz_2}(a), we plot Re$\chi_{xxx}^{(2)}(\omega_t, \omega_\tau)$ in strongly confined regime with $h_z/J=0.4$. Unlike the previous regime, it contains a non-rephasing like signal which appears as strong cross (off-diagonal) peaks in the first quadrant. Besides, a subdominant rephasing like signal appears as cross peaks in the fourth quadrant.

To interpret the 2D spectra, we use a projected TK model and can study the excitations of Eq.(\ref{eq:tfim_l}) perturbatively~\cite{rutkevich2010weak,kormos2017real}.  The idea is to project the full Hilbert space down to the Hilbert space of TK states, where regions of opposite magnetization are separated by the two different domain-walls. In this model, each TK state is represented as $\ket{j, l} \equiv \ket{...\uparrow\uparrow \downarrow_j\downarrow...\downarrow\downarrow_{(j+l-1)} \uparrow\uparrow...}$, where $j$ is the starting site of down spins. The TK model has been adopted to phenomenologically understand the confinement of excitations observed in the dynamical (1D) neutron response of $\mathrm{CoNb}_2\mathrm{O}_6$~\cite{coldea2010quantum,rutkevich2010weak}. The projected model is expected to capture the low-energy excitations of the original model well as long as $h_x/J \ll h_z/J$. This can be understood by looking into the energy gap between TK states and four-kink states~\cite{verdel2020real}. The TK Hamiltonian, $\mathcal{H}_{TK} \equiv \mathcal{P} H \mathcal{P}$ with the projector $\mathcal{P}$, acts as follows:
\bea
\nonumber
\mathcal{H}_{TK}\ket{j, l}&=& 4J\ket{j, l} - h_x \big[\ket{j, l+1}+ \ket{j, l-1} 
\\ 
&+&\ket{j+1, l-1}+ \ket{j-1, l+1} \big] + 2h_z l \ket{j, l}.
\nonumber
\\
\label{eq:H_tk_r}
\eea
For our translational invariant model, the total momentum $p$ of the bound state is a good quantum number. In the momentum basis $\ket{p,l}=\sum_{j}\exp(i p j)\ket{j,l}$, the Hamiltonian is diagonal in $p$ and acts on $\ket{p, l}$ as
\bea
\nonumber
\mathcal{H}_{TK} \ket{p,l} &=& 4J \ket{p,l}-h_x \big[ (1+e^{\text{i} p}) \ket{p,l+1}
\\
&+&(1+e^{-\text{i} p})\ket{p,l-1} \big] + 2h_z l \ket{p,l}.
\label{eq:H_tk}
\eea

By solving the eigen-equation in Eq.~(\ref{eq:H_tk}) for a given momenta $p$ and a band index $n$, one get TK bound state $\ket{\Phi_n(p)}$ with the excitation energy $E_n(p)$, which is shown in the inset of Fig.~\ref{fig:hz_1}(b) and \ref{fig:hz_2}(b). (See Section III of SI for details). Then, following Ref.~\onlinecite{rutkevich2010weak}, which calculates the linear response, we find (after some algebraic manipulation)
\bea
\chi_{xxx}^{(2)}(t,\tau + t ) &=& \theta(t)\theta(\tau) (Q_1 + Q_2 + Q_3 + Q_4)
\eea 
with
\bea
\nonumber
Q_1 &=& \sum_{n,p} C_{n,n}(p)\cos \big[ E_n(p) (t + \tau) \big],
\\
\nonumber
Q_2 &=& \sum_{n,p}C_{n,n}(p)\cos \big[ E_n(p) \tau \big],
\\
\nonumber
Q_3 &=& \sum_{n,m,p}C_{n,m}(p)\cos \big[ E_n(p) t + E_m(p) \tau \big],
\\
\nonumber
Q_4 &=& \sum_{n,m,p}C_{n,m}(p)\cos \big[ \big(E_n(p)-E_m(p)\big) t - E_m(p)\tau \big]
\eea
where $C_{n,m}(p)$ is the optical matrix element which depends on band indices $n, m$ and momentum $p$ (See Section III of SI for details). 

The interpretation of 2D spectra now becomes transparent: $Q_1$ gives rise to diagonal non-rephasing peaks at $\omega_t, \omega_{\tau}=E_n(p)$ and $Q_2$ produces dominant terahertz rectification signals at $\omega_t=0$ and $\omega_{\tau}=\pm E_m(p)$ [Fig.~\ref{fig:hz_1}(b) and Fig.~\ref{fig:hz_2}(b)]. In the strongly confined regime, $Q_3$ gives rise to dominant cross peaks in the first frequency quadrant originating from the non-rephasing like process [Fig.~\ref{fig:hz_2}(b)]. To be more precise, such peaks sharply appear at $\omega_t\!=\!E_1(0)$ and $\omega_{\tau}\!=\!E_2(0)$ or $\omega_t\!=\!E_2(0)$ and $\omega_{\tau}\!=\!E_1(0)$, indicating the presence of multiple TK excited states. Such sharp peaks do not appear in the free 1D TFIM, which can be mapped to an independent two-level systems with each having a single excited state~\cite{wan2019resolving}. $Q_4$ contains terms which induce sharp (non-)rephasing like signals in the first (fourth) frequency quadrant which are visible in the strongly confined regime [Fig.~\ref{fig:hz_2}(b)]. Such signals also originate from the presence of multiple excited states and appear at $\omega_t=E_2(0)-E_1(0)$, an energy gap between first and second excited states, and $\omega_{\tau}=E_2(0)$ or $-E_1(0)$, see inset.
	
{\em Conclusions.}
In the present work, we have developed an iMPS method for calculating the nonlinear response of 1D spin systems. As a demonstration, we calculated the second order susceptibility, which dominates the nonlinear response, for the ferromagnetic 1D TFIM where a single spin flip is fractionalized into two freely moving domain-walls. We benchmarked our numerical results with exact analytical calculations. We then included a longitudinal field, which induces a linear confining potential between kink excitations. In the presence of a longitudial field, the second order susceptibility contains new signals which give rise to strong non-rephasing and rephasing like peaks. To understand the emergence of such signals, we employ a simplified two-kink description, which describes the low-energy excitations in the strongly confined regime, and calculate the second order susceptibility perturbatively. The approximate method captures the nonlinear response of the system in the strongly confined regime and allows for a simple interpretation.

As a future direction, it would be interesting to apply our iMPS method near the quantum critical point between the ferromagnetic and paramagnetic states where a hidden E$_8$ symmetry emerges \cite{coldea2010quantum,kjall2011bound}. A crucial question regards then the existence of robust signals in 2D spectrum which detect the emergent symmetry. Regarding the microscopic description and 3D nature of  $\textrm{CoNb}_2\textrm{O}_6$~\cite{coldea2010quantum,morris2014hierarchy} ($\textrm{BaCo}_2\textrm{V}_2\textrm{O}_8$~\cite{zou20218,faure2018topological}) it would be interesting to study a more quantitative model~\cite{fava2020glide,morris2021duality} and go beyond simple chains, e.g. by extending these to coupled Ising ladders~\cite{morris2014hierarchy}. Terahertz 2D coherent spectroscopy holds the promise of uncovering the nature of exotic excitations in strongly correlated quantum materials. A challenging but very worthwhile direction will be an extension of our method to quantum magnets beyond one-dimension where the nature of fractionalized excitations and confinement thereof remains poorly understood.

{\em Note added.}
When finalizing the manuscript, related works appeared that investigate nonlinear response from quasiparticle interactions \cite{fava2022divergent, hart2022self}.

\section{Acknowledgments} We thank N. P. Armitage, R. Coldea, M. Drescher, H.-K. Jin, and W. Choi for insightful discussions related to this work. G.B.S. is funded by the European Research Council (ERC) under the European Unions Horizon 2020 research and innovation program (grant agreement No. 771537). F.P. acknowledges the support of the Deutsche Forschungsgemeinschaft (DFG, German Research Foundation) under Germany’s Excellence Strategy EXC-2111-390814868. J. K. acknowledges support from the Imperial-TUM flagship partnership. The research is part of the Munich Quantum Valley, which is supported by the Bavarian state government with funds from the Hightech Agenda Bayern Plus. Tensor network calculations were performed using the TeNPy Library \cite{hauschild2018efficient}.

\bibliographystyle{apsrev4-2}
\bibliography{non_tfim_bib}
\renewcommand{\thefigure}{S\arabic{figure}}
\setcounter{figure}{0}
\renewcommand{\theequation}{S\arabic{equation}}
\setcounter{equation}{0}

\begin{widetext}

\section{Supplementary Material}

\subsection{A procedure to obtain $S^{\beta\gamma\alpha}_{j,l,c}(\tau,\tau+t,0)$ with iMPS method}
In this section, we provide steps to calculate $S^{\beta\gamma\alpha}_{j,l,c}(\tau,\tau+t,0)= e^{iE\tau} \bra \psi \sigma^\beta_j e^{iHt} \sigma^\gamma_l e^{-iH(\tau + t)} \sigma^\alpha_{c} \ket \psi$ using iMPS.
\label{sec:IMPS}
	\begin{enumerate}
	\item
	Find an iMPS approximation of the ground state $\ket{\psi}$ with energy $E$ \cite{mcculloch2008infinite,schollwock2011density}.
	\item
	Allow the tensors of the iMPS with a spatial window of size $L$ to vary in time as in Refs.~\onlinecite{phien2012infinite,milsted2013variational,binder2018infinite,ejima2021finite}.
	\item
	Apply a local operator $\sigma^\beta_{c}$ ($\sigma^\alpha_{c}$) at the center of the window to get $\sigma^\beta_{c}\ket{\psi}$ ($\sigma^\alpha_{c}\ket{\psi}$).
	\item
	Perform a real time evolution following the local quench $\sigma^\beta_{c}$ ($\sigma^\alpha_{c}$) using iTEBD method~\cite{vidal2003efficient,vidal2004efficient,vidal2007classical} to obtain an iMPS which represents $e^{-iHt}\sigma^\beta_{c}\ket{\psi}$ ($e^{-iH(\tau + t)} \sigma^\alpha_{c} \ket \psi$).
	\item
	Shift every $B$ tensor of an iMPS, which represents $e^{-i H t}\sigma^\beta_{c}\ket{\psi}$, $(c-j)$ sites to the right within the window to obtain a new iMPS ``bra" associated to $e^{-iHt}\sigma^\beta_{j}\ket{\psi}$.
	\item
	Apply an operator $\sigma^\gamma_{l}$ to an iMPS, which approximates $e^{-iH(\tau + t)} \sigma^\alpha_{c} \ket \psi$, and get a new iMPS ``ket" associated to $\sigma^\gamma_l e^{-iH(\tau + t)} \sigma^\alpha_{c} \ket \psi$.
	\item
	Evaluate an overlap of two iMPS ``bra" and ``ket" within the window by calculating the dominant eigenvalue of the corresponding transfer matrix to obtain $\bra \psi \sigma^\beta_j e^{iHt} \sigma^\gamma_l e^{-iH(\tau + t)} \sigma^\alpha_{c} \ket \psi$.
	\item
	Multiply $e^{iE\tau}$ and $\bra \psi \sigma^\beta_j e^{iHt} \sigma^\gamma_l e^{-iH(\tau + t)} \sigma^\alpha_{c} \ket \psi$.
\end{enumerate}
 
\subsection{Imaginary part of $\chi^{(2)}_{xxx}(\omega_t, \omega_\tau)$}
\label{sec:imag}

In this section, we focus on Im$\chi_{xxx}^{(2)}(\omega_t, \omega_\tau)$ in the ferromagnetic phase of the Ising model. In Fig.~\ref{fig:chi2_im}(a), we plot Im$\chi_{xxx}^{(2)}(\omega_t, \omega_\tau)$ for $h_x/J=0.2$ without longitudinal field using iMPS method. To obtain the result, we directly perform fast Fourier transformations of the real time correlation functions $\chi^{(2)}_{xxx}(t,\tau+t)$. Fig.~\ref{fig:chi2_im}(b) shows Im$\chi^{(2)}_{xxx}(\omega_t, \omega_\tau)$ in weakly confined regime with $h_z/J=0.03$. One can clearly observe the emergence of strong non-rephasing like signals in the first frequency quadrant. In Fig.~\ref{fig:chi2_im}(c), we plot Im$\chi^{(2)}_{xxx}(\omega_t, \omega_\tau)$ in strongly confined regime with $h_z/J=0.4$. Simliar to the real part, it contains new signals which include dominant non-rephasing and strong rephasing like peaks.

\begin{figure}[h!]
	\begin{minipage}{0.28\linewidth}
		\vspace*{0.1cm}
		\subfloat[Im$\chi_{xxx}^{(2)}(\omega_t, \omega_\tau)$ without longitudial field]{\label{fig:chi2_tfim_im}\includegraphics[scale=0.4]{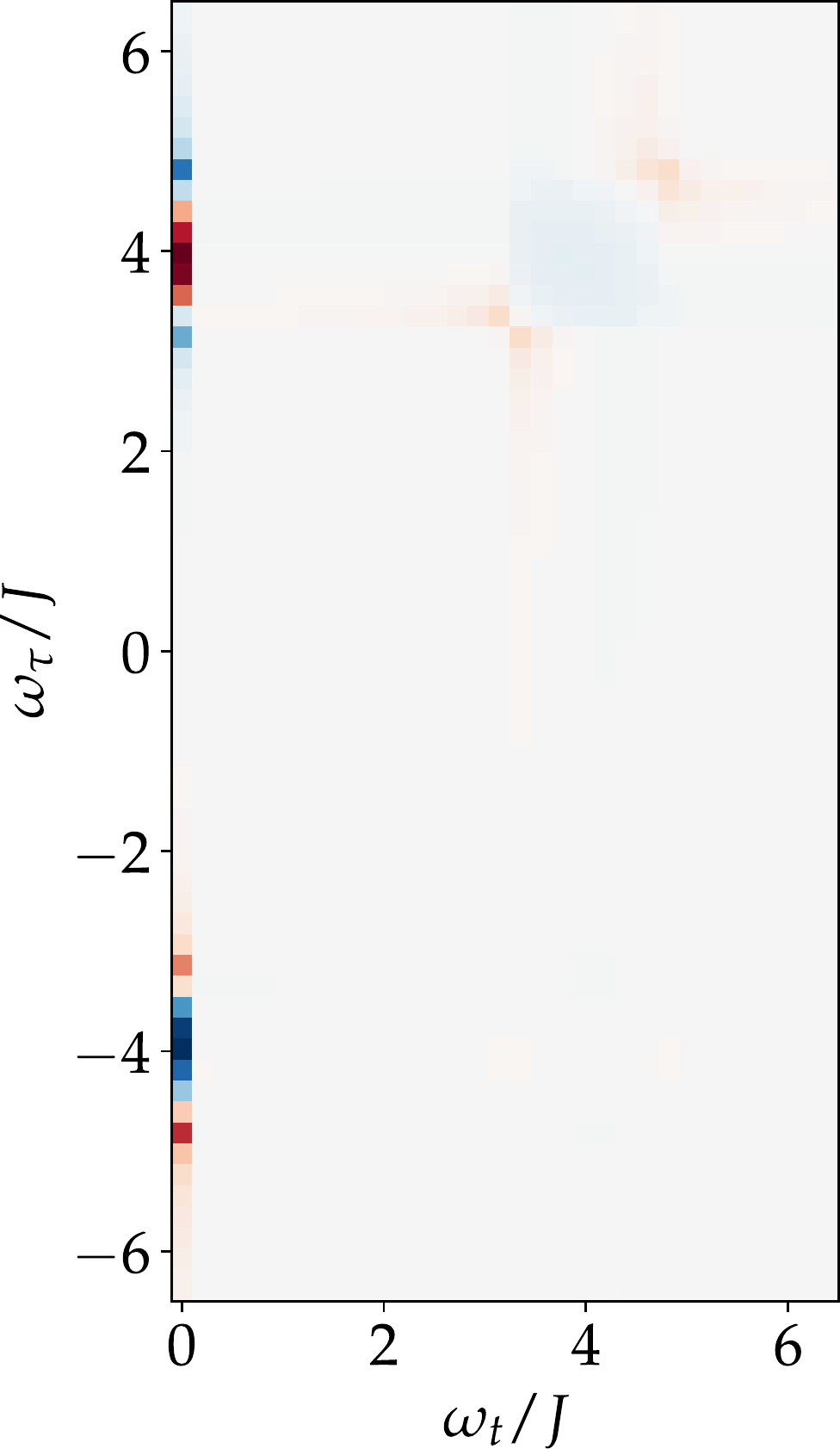}}
	\end{minipage}%
	\begin{minipage}{0.28\linewidth}
		\subfloat[Im$\chi_{xxx}^{(2)}(\omega_t, \omega_\tau)$ in weakly confined regime]{\label{fig:chi2_1_im}\includegraphics[scale=0.4]{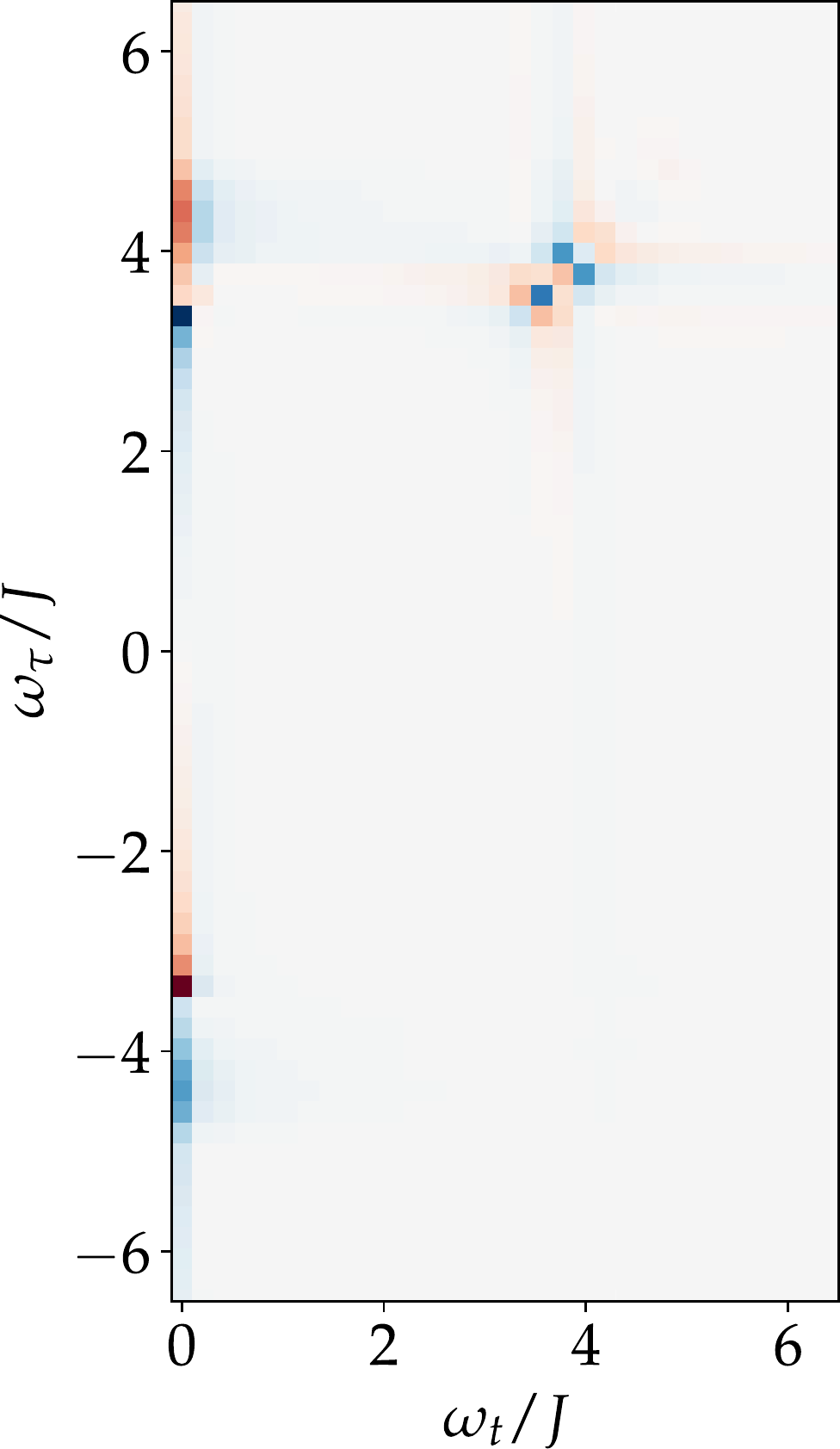}}
	\end{minipage}%
	\begin{minipage}{0.28\linewidth}
		\subfloat[Im$\chi_{xxx}^{(2)}(\omega_t, \omega_\tau)$ in strongly confined regime]{\label{fig:chi2_2_im}\includegraphics[scale=0.4]{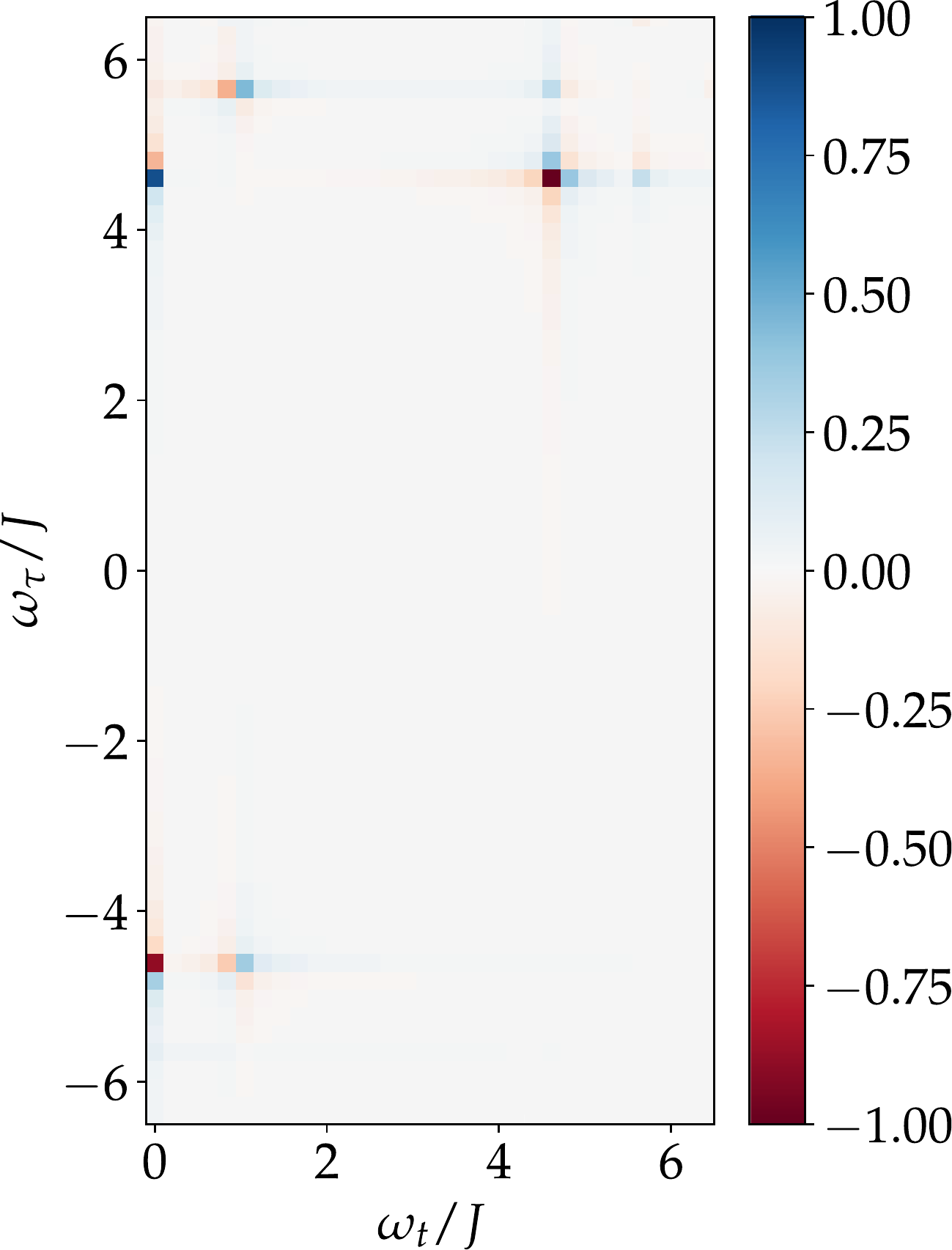}}
	\end{minipage}%
	\caption{(color online)} 
	\label{fig:chi2_im}
\end{figure}
 
\subsection{Three point spin correlation functions within TK model}
\label{sec:tsc}

In this section, we follow Ref.~\onlinecite{rutkevich2010weak}, which calculates the two point spin correlator, and formulate the three point correlator within the projected TK low energy subspace. In this subspace, each TK state is represented as $\ket{j, l} \equiv \ket{...\uparrow\uparrow \downarrow_j\downarrow...\downarrow\downarrow_{(j+l-1)} \uparrow\uparrow...}$, where $j$ is the starting site of down spins. The TK Hamiltonian $\mathcal{H}_{TK}$ acts as follows:
\bea
\mathcal{H}_{TK}\ket{j, l}&=& 4J\ket{j, l} - h_x \big[\ket{j, l+1}+ \ket{j, l-1} +\ket{j+1, l-1}+ \ket{j-1, l+1} \big] + 2h_z l \ket{j, l}.
\label{eq:s_H_tk_r}
\eea
In the momentum basis $\ket{p,l}=\sum_{j}\exp(i p j)\ket{j,l}$, the Hamiltonian is diagonal in $p$ and acts on $\ket{p, l}$ as
\bea
\mathcal{H}_{TK} \ket{p,l} &=& 4J \ket{p,l}-h_x \big[ (1+e^{\text{i} p}) \ket{p,l+1}+(1+e^{-\text{i} p})\ket{p,l-1} \big] + 2h_z l \ket{p,l}.
\label{eq:s_H_tk}
\eea
Now, the eigen-equation with the excitation energy $E_n(p)$,
\bea
\mathcal{H}_{TK} |\Phi_n(p)\rangle = E_n(p) |\Phi_n(p)\rangle  \label{ei}
\eea
where $|\Phi_n(p)\rangle \equiv \sum_{l=1}^{\infty} \exp(\frac{\text{i} p l}{2}) \psi_n(l,p) \ket{p,l} / \sqrt{\sum_{l=1}^{\infty}|\psi_n(l,p)|^2}$ with the site integer $l$, momentum $p$, and discrete band index $n$ takes the form as below.
\bea
[4 J +2 h_z l - E_n(p)] \psi_n(l,p) -2h_x\cos(p/2)[ \psi_n(l+1,p)+\psi_n(l-1,p)]=0
\label{eq:s_eig}
\eea
We rewrite Eq.~(\ref{eq:s_eig}) with dimensionless parameters as follows
\bea
\left(-\lambda_n+\mu \,l\right) \psi_n(l,p) 
-\frac{\psi_n(l+1,p)+\psi_n(l-1,p)}{2}=0
\label{eq:s_eig_2}
\eea
where
\bea
\lambda_n \equiv \frac{E_n(p)-4 J}{4 h_x \cos(p/2)} , \quad \mu \equiv \frac{h_z}{2h_x \cos(p/2)}
\label{eq:s_para}
\eea
with the boundary conditions, $\lim_{l\to0}\psi_n(l,p)=0$ and $\lim_{l\to+\infty}\psi_n(l,p)=0$ \cite{rutkevich2010weak}. The result for the eigenvalues $\lambda_n$ reads as
\bea
\lambda_n=-\mu \,\nu_n
\eea
where $\nu_n$ are the solutions of the equation
\bea
J_{\nu_n}(1/\mu)=0.
\label{eq:s_sol}
\eea
Here, $J_{\nu}(x)$ is the Bessel function of order $\nu$. Then, the solution of the Eq.~(\ref{eq:s_eig}) reads as
\bea
E_n(p) =4 J-2h_z\,\nu_n. 
\label{eq:s_energy}
\eea

As given in the main text, $\chi^{(2)}_{xxx}(t,\tau+t)$ is written as 
\bea
\chi^{(2)}_{xxx}(t,\tau+t) = &-& \frac{\theta(t)\theta(\tau)}{4L} \mathrm{Re}\sum_{j,k,l}\big[ S^{xxx}_{j,k,l}(\tau+t,\tau,0)- S^{xxx}_{j,k,l}(\tau,\tau+t,0) \big]
\label{eq:s_chi2}
\eea
where $S^{xxx}_{j,k,l}(s_1,s_2,s_3) \equiv \langle \sigma^x_j(s_1) \sigma^x_k(s_2) \sigma^x_l (s_3)\rangle$ is the three point correlation function. The first term of Eq.~(\ref{eq:s_chi2}) can be rewritten as follows :
\bea
\nonumber
\sum_{j,k,l} S^{xxx}_{j,k,l}(\tau+t,\tau,0)&=&\sum_{j,k,l} \langle \psi | e^{\text{i}H(\tau+t)} \sigma^x_j e^{-\text{i} H t} \sigma^x_k e^{-\text{i} H \tau} \sigma^x_l |\psi \rangle \\
\nonumber
&=& \sum_{j,k,l}\sum_{n,m,p,q} e^{-\text{i} (E_n(p) t + E_m(q) \tau )} \langle \psi | \sigma^x_j | \Phi_n(p) \rangle \langle \Phi_n(p) | \sigma^x_k | \Phi_m(q) \rangle \langle \Phi_m(q) | \sigma^x_l |\psi \rangle \\
&=& \sum_{j,k,l}\sum_{n,m,p} e^{-\text{i} (E_n(p) t + E_m(p) \tau)} \langle \psi | \sigma^x_j | \Phi_n(p) \rangle \langle \Phi_n(p) | \sigma^x_k | \Phi_m(p) \rangle \langle \Phi_m(p) | \sigma^x_l |\psi \rangle
\label{eq:s1}
\eea
where $\ket{\psi}$ is the non-degenerate ferromagnetic ground state with no kinks.
In Eq.~(\ref{eq:s1}), $\langle \Phi_n(p) | \sigma^x_k | \Phi_m(p) \rangle$ can be expanded as
\bea
\sum_k \langle \Phi_n(p) | \sigma^x_k | \Phi_m(p) \rangle = \frac{\sum_k \sum_{l',l}\langle p, l'|\sigma^x_k|p, l\rangle e^{-\text{i} p (l'-l)/2} \psi_n(l',p)\psi_m(l,p)} {\sqrt{\sum_{l=1}^{\infty}|\psi_n(l,p)|^2} \sqrt{\sum_{l=1}^{\infty}|\psi_m(l,p)|^2}}
\label{eq:s2}
\eea
with
\bea
\langle p, l'|\sigma^x_k|p, l\rangle = \sum_{j',j}\exp(-\text{i} p (j'-j)) \langle j', l'|\sigma^x_k|j,l\rangle.
\label{eq:s2_2}
\eea
Eq.~(\ref{eq:s2_2}) clearly shows that Eq.~(\ref{eq:s2}) is finite only when one of the following four conditions is met.
\begin{alignat*}{2}
\nonumber
&1.~|j,l\rangle = |k+1,l\rangle,~~ &&|j',l'\rangle = |k,l+1\rangle \\
\nonumber
&2.~|j,l\rangle = |k-l,l\rangle,~~ &&|j',l'\rangle = |k-l,l+1\rangle, \\
\nonumber
&3.~|j,l\rangle = |k,l\rangle,~~ &&|j',l'\rangle = |k+1,l-1\rangle, \\
&4.~|j,l\rangle = |k-l+1,l\rangle,~~ &&|j',l'\rangle = |k-l+1,l-1\rangle 
\label{eq:s3}
\end{alignat*}
Then, Eq.~(\ref{eq:s2}) can be simplified as the following.
\bea
\sum_k \langle \Phi_n(p) | \sigma^x_k | \Phi_m(p) \rangle = \frac{\sum_{l} \cos{\frac{p}{2}} \big[ \psi_n(l,p) \psi_m(l+1,p)+ \psi_n(l,p)  \psi_m(l-1,p) \big]}{\sqrt{\sum_{l=1}^{\infty}|\psi_n(l,p)|^2} \sqrt{\sum_{l=1}^{\infty}|\psi_m(l,p)|^2}}
\label{eq:s4}
\eea
To formulate Eq.~(\ref{eq:s4}), we first rewrite Eq.~(\ref{eq:s_eig_2}) with two different band indices, $n$ and $m$.
\bea
\label{eq:s5}
&&(-\lambda_n+\mu \,l) \psi_n(l,p) = \frac{\psi_n(l+1,p)+\psi_n(l-1,p)}{2}\\
\label{eq:s6}
&&(-\lambda_m+\mu \,l) \psi_m(l,p) = \frac{\psi_m(l+1,p)+\psi_m(l-1,p)}{2}
\eea
Now, we multiply $\psi_m(l+1,p)$ to Eq.~(\ref{eq:s5}) and $\psi_n(l-1,p)$ to Eq.~(\ref{eq:s6}), subtract one from the other, and sum it over $n$. Then, we get an expression for $\sum_l \psi_n(l,p) \psi_m(l+1,p)$ :
\bea
\sum_l \psi_n(l,p) \psi_m(l+1,p) = \frac{\psi_n(0,p)\psi_m(2,p)-\psi_n(1,p)\psi_m(1,p)}{2(\lambda_n-\lambda_m-\mu)}  = \frac{-2}{\lambda_n-\lambda_m-\mu}
\eea
where we set $\psi_n(1,p)=-2$ without loss of generality \cite{rutkevich2010weak}.
We can proceed similarly and get
\bea
\sum_l \psi_n(l,p) \psi_m(l-1,p) = \frac{-2}{\lambda_m-\lambda_n-\mu}
\eea
In the end, we get expressions for $\sum_{j,k,l} S^{xxx}_{jkl}(\tau+t,\tau,0)$ as the following.
\bea
&&\sum_{j,k,l} S^{xxx}_{jkl}(\tau+t,\tau,0) = \sum_{n,m,p} C_{n,m}(p) e^{-\text{i} (E_n(p) t + E_m(p) \tau)}
\label{eq:s7}
\eea
with the optical matrix element
\bea
C_{n,m}(p) \equiv -2\cos{\frac{p}{2}} (\frac{1}{\lambda_m-\lambda_n-\mu} +\frac{1}{\lambda_n-\lambda_m-\mu} ) I_{n}(p) I_{m}(p).
\eea
Here, the relative intensity of the $n$-th mode is defined as
\bea
I_{n}(p) \equiv \frac{|\psi_n(1,p)|^2}{\sum_{l=1}^{\infty}|\psi_n(l,p)|^2} = \frac{4}{\sum_{l=1}^{\infty}|\psi_n(l,p)|^2} = 2\mu \left\{\frac{\partial}{\partial \nu}\left[\frac{J_\nu(1/\mu)}{J_{\nu+1}(1/\mu)}\right]\right\}^{-1} \bigg|_{\nu\to\nu_n}
\eea
where $\nu_n$ is the $n$-th solution of Eq.~(\ref{eq:s_sol}) \cite{rutkevich2010weak}. Similiar formulation gives
\bea
\sum_{j,k,l} S^{xxx}_{jkl}(\tau,\tau+t,0) = \sum_{n,m,p} C_{n,m}(p) e^{-\text{i} E_m(p) \tau} e^{-\text{i} (E_m(p)-E_n(p)) t}.
\eea

\end{widetext}

\end{document}